\documentclass[twocolumn,showpacs,superscriptaddress,floatfix,prb]{revtex4}
\usepackage{graphicx,amsfonts,amssymb,amsmath,hyperref,hypcap}

\newcommand{\Tr}{\mathop{\mathrm{Tr}}\nolimits}

\hypersetup{
  colorlinks = {true}
}

\begin{document}


\title{Phase diagram of the toric code model in a parallel magnetic field}

\author{Fengcheng Wu}
\affiliation{Hefei National Laboratory for Physical Sciences at Microscale and Department of Modern Physics, University of Science and Technology of China, Hefei, Anhui 230026, China}

\author{Youjin Deng}
\email{yjdeng@ustc.edu.cn}
\affiliation{Hefei National Laboratory for Physical Sciences at Microscale and Department of Modern Physics, University of Science and Technology of China, Hefei, Anhui 230026, China}
\affiliation{Department of Physics, University of Massachusetts, Amherst, Massachusetts 01003, USA}

\author{Nikolay Prokof'ev}
\email{prokofev@physics.umass.edu}
\affiliation{Department of Physics, University of Massachusetts, Amherst, Massachusetts 01003, USA}


\begin{abstract}
Ground-state phase diagram of the toric code model in a parallel magnetic field has three distinct
phases: topological, charge-condensed, and vortex-condensed states.
To study it we consider an implicit local order parameter characterizing the
transition between the topological and charge-condensed phases, and sample it using
continuous-time Monte Carlo simulations. The corresponding second-order
transition line is obtained by finite-size scaling analysis of this order parameter.
Symmetry breaking between charges and vortices along the first-order transition
line is also observed, and our numerical result shows that the end point of the first-order transition line is
located at $h_x^{(c)}=h_z^{(c)}=0.418(2)$.
\end{abstract}


\maketitle

\section{Introduction}
Topological phases are attracting a lot of attention for a variety of reasons from
precise determination of physical constants to unusual properties of quasiparticles
and promise for quantum information and computation applications. The search for such phases has been mainly focused on electron systems, such as quantum Hall states and topological insulators\cite{Hasan10, Qi11}. However, topological phases can also exist in magnetic
systems and the toric code model (TCM)\cite{Kitaev03} based on spin-1/2 degrees of freedom
provides a key example. TCM can be viewed as the anisotropic limit of Kitaev's honeycomb model\cite{Kitaev06}, which has recently been suggested to emerge in iridium oxides\cite{Jackeli09, Chaloupka10, Choi12}.

The ground state of TCM in zero external field is topologically ordered and characterized by two types
of bosonic excitations (charges and vortices) with mutual semionic statistics. To be useful for quantum computing, the topological character of TCM should be robust against local magnetic field perturbations. Thus one of the most
important questions to ask is what are the critical values of magnetic fields which
cause condensation of charges or vortices, i.e. break the topological order.
The purpose of this paper is to perform an accurate study of phase transition lines
in TCM subject to a parallel magnetic field.

As TCM involves four-spin interactions, the continuous-time Monte Carlo approach of Ref.~\onlinecite{Prokof'ev98}
has to be modified appropriately and supplemented with additional updates. The topological phase is characterized by non-local quantum correlations.
However, an implicit local order parameter can still be defined to characterize the transition between the topological phase and charge-condensed phase.
The corresponding second-order transition line is then obtained from the
standard finite-size scaling analysis of this order parameter. Along the self-duality line, there is a symmetry between charges and vortices which is, however, broken
in a certain range of parameters. It is observed in the Monte Carlo simulation
as the first-order phase transition between the charge- and vortex-condensed phases.

Similar studies of TCM were performed in the past using other methods \cite{Tupitsyn10,Vidal09, Vidal09_2, Dusuel11}.
In Ref.~\onlinecite{Tupitsyn10} the authors used an approximate mapping of TCM onto the anisotropic $\mathbb{Z}_{2}$ gauge Higgs model,
and computed the phase diagram of the corresponding 3D classical model
in the isotropic limit, extending previous studies of the
$\mathbb{Z}_{2}$ gauge Higgs model, see Ref.~\onlinecite{Genovese}, to larger system sizes.
The behavior of TCM in a parallel field and/or a transverse field
was also studied using several perturbative (high order) treatments \cite{Vidal09, Vidal09_2, Dusuel11}.
The Monte Carlo algorithm developed in this paper suffers from the
sign-problem in the presence of the transverse field component.
Therefore we restrict our studies to the parallel field in which case
the parameter space of the model considered here is the same as in Ref.~\onlinecite{Vidal09}.
Not surprisingly, we find the same topology of the phase diagram,
but our numerical result shows that the end point of the first-order transition line is located at $h_x^{(c)}=h_z^{(c)}=0.418(2)$,
which differs from $h_x^{(c)}=h_z^{(c)}=0.48(2)$ calculated in Ref.~\onlinecite{Vidal09}.
To elaborately compare our Monte Carlo data with the series expansion results,
we also compute two magnetization-like quantities using the bare series expansion
given in Ref.~\onlinecite{Vidal09} and \onlinecite{Dusuel11}.
We argue that the discrepancy in the end point is mainly because the series expansion about high-field limit at order 5
is questionable in the field strength of interest.

\section{Model}
The toric code model is defined on a square lattice with spins located on lattice bonds.
The Hamiltonian (in zero field) is based on four-spin interactions:
\begin{equation}
H_{TCM} = -J_s \sum_s A_s - J_p \sum_p B_p,
\label{H_TCM}
\end{equation}
\begin{figure}[htbp]
\vspace{0cm}
 \includegraphics[width=0.70\columnwidth]{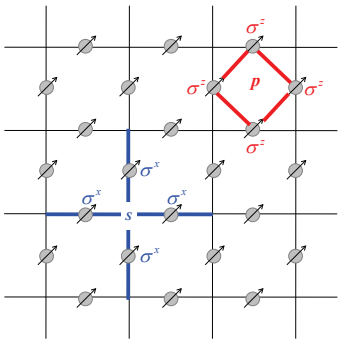}
 \caption{(Color online) An illustration of the toric code model geometry. $A_s = \prod_{j \in s} \sigma^{x}_j$ and $B_p = \prod_{j \in p} \sigma^{z}_j$
 are products of spin operators on bonds incident to site $s$ and surrounding plaquette $p$, respectively.}
 \label{tcm}
\end{figure}
where $A_s = \prod_{j \in s} \sigma^{x}_j$ and $B_p = \prod_{j \in p} \sigma^{z}_j$ ($\sigma_{j}^{\alpha}$ are the Pauli matrices). Here subscripts
$s$ and $p$ refer to sites and plaquettes of the square lattice, see Fig.~\ref{tcm}, respectively, and address all spins surrounding the corresponding site or plaquette.
Since all $A_s$ and $B_p$ commute with each other, the ground state manifold
is known exactly. It corresponds to eigenstates of all $A_s$ and $B_p$ with
maximal eigenvalues +1. The ground-state degeneracy depends on boundary conditions,
and on a torus there are four degenerate ground states.
Elementary excitations, a charge on site $s$ or a vortex on plaquette $p$, correspond to states with eigenvalues $A_s=-1$ or $B_p=-1$. When a charge
(vortex) is circulated around another vortex (charge) the wave function acquires
a phase shift of $\pi$. Thus charges and vortices have mutual semionic statistics. On the other hand, composite charge-vortex particles have fermionic statistics.

In the presence of a parallel magnetic field, which is perpendicular to the $y-$direction,
the Hamiltonian takes the form:
\begin{equation}
H = H_{TCM} - h_x \sum_b \sigma^x_b - h_z \sum_b \sigma^z_b,
\label{H}
\end{equation}
where $b$ refers to lattice bonds. By orienting individual spins the
$h_z$ and $h_x$ terms reduce energy gaps for creating charge and vortex excitations,  respectively,
and ultimately cause their condensation.

\section{Continuous-Time Monte Carlo Simulation}
To achieve complete quantum mechanical solution of the model we use the
continuous-time Monte Carlo approach introduced in Ref.~\onlinecite{Prokof'ev98}.
Within this approach, the Hamiltonian(\ref{H}) is split into two non-commuting
terms $H=U+K$ with \begin{equation}
\begin{aligned}
U=-J_p \sum_{p}B_{p}-h_z \sum_b \sigma^z_b,\\[2pt]
K= -J_s \sum_{s}A_{s}-h_x \sum_b \sigma^x_b.
\label{UK}
\end{aligned}
\end{equation}
In the basis of $\sigma^z_b$ eigenstates, $U$ and $K$ are conveniently called
potential and kinetic energy terms, respectively. Similarly to the standard
Feynman's lattice path integral in the interaction representation we
expand the partition function $Z$ in powers of $K$ as follows:
\begin{eqnarray}
&Z=\Tr[\exp(-\beta H)]&\nonumber
\\&=\sum\limits_{n=0}^{\infty}\int_{0<\tau_1<\tau_2<\cdots<\tau_n}^{\beta}d\tau_1\cdots d\tau_n\sum_{\{\alpha_{0},\alpha_{1} \cdots \alpha_{n-1},\alpha_{n}=\alpha_{0}\}}&\nonumber
\\&(-1)^{n}K_{\alpha_{n}\alpha_{n-1}}\cdots K_{\alpha_{2}\alpha_{1}}K_{\alpha_{1}\alpha_{0}}\exp\{-\int_{0}^{\beta}U(\tau)d\tau\}.&
\label{Z}
\end{eqnarray}
Here $\alpha_j$ with $j=1,2\cdots n$ enumerate $\sigma_b^z$ basis states
characterized by a set of classical variables $z_b=\pm 1$ on all bonds;
$\tau_j$ 
are the imaginary time moments when the state $\alpha_{j-1}$ is changed to
$\alpha_j$ due to action of the kinetic energy term,
$K_{\alpha_{j}\alpha_{j-1}}=\langle \alpha_{j} \mid K\mid\alpha_{j-1} \rangle$.
By definition, the four-spin operator $A_s$ simultaneously flips four spins on the bonds connected to the site $s$, while operator $\sigma_b^x$ flips only the spin on bond $b$.
Finally, the  dependence of potential energy on time in the exponential
of (\ref{Z}) is defined as
$ U(\tau) = U_{\alpha_i}$ for $\tau \in (\tau_i, \tau_{i+1})$.
Note that the weight for each configuration is positive here, while the inclusion of the transverse field component will result in the sign problem.

\begin{figure}[htbp]
\vspace{0cm}
 \includegraphics[width=0.80\columnwidth]{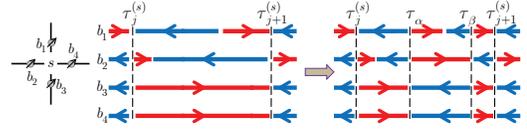}
 \caption{(Color online) An illustration of creating a pair of four-spin flips.}
 \label{four_spin_flip_creation}
\end{figure}

\begin{figure}[htbp]
\vspace{0cm}
 \includegraphics[width=0.80\columnwidth]{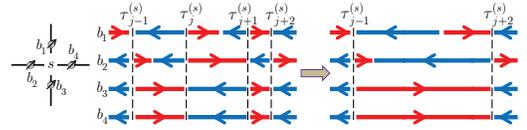}
 \caption{(Color online) An illustration of annihilating a pair of four-spin flips.}
 \label{four_spin_flip_annihilation}
\end{figure}

\begin{figure}[htbp]
\vspace{0cm}
 \includegraphics[width=0.8\columnwidth]{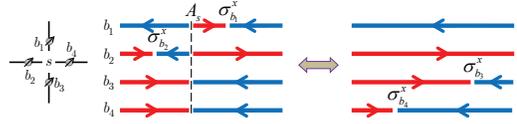}
 \caption{(Color online) An illustration of \textbf{Update 5}.
 From left to right, the update annihilates a four-spin flip on site $s$ and two one-spin flips on bonds $b_1$ and $b_2$
 and creates two one-spin flips on bonds $b_3$ and $b_4$.}
 \label{kink}
\end{figure}

The continuous-time Monte Carlo has been mainly used for quantum
models with two-body interactions,
such as the Bose-Hubbard model, where the worm algorithm \cite{Prokof'ev98}
provides an effective approach to large-scale and high-efficiency simulations.
Here we apply the method to quantum
models with four-spin non-diagonal interactions. An obvious set of elementary
Monte Carlo updates, which
(1) create/annihilate a pair of one-spin flips,
(2) change times of one-spin flips,
(3) create/annihilate a pair of four-spin flips, and
(4) change times of four-spin flips,
is not ergodic for Hamiltonian (\ref{H}) because it does not admit all allowed
configurations with alternating single-spin and four-spin flips.
This problem is taken care of by an update which involves both four-spin and one-spin flips. In the following, we provide a detailed description of all updates.
\begin{figure}[htbp]
\vspace{0cm}
\includegraphics[width=0.85\columnwidth]{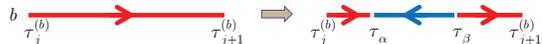}
\caption{(Color online) Create a pair of one-spin flips. }
 \label{one_spin_flip_creation}
\end{figure}
\\[2ex]\textbf{Update 1} Create/Annihilate a pair of one-spin flips.
\\\textbf{Step} 1.1 Select an bond $b$ at random.
Let $N(b)$ denote the number of one-spin flips on bond $b$.
\\\textbf{Step} 1.2 Keep $\sigma_b^z (\tau=0) $ unchanged, and
propose to create or annihilate a pair of one-spin flips on bond $b$ with equal probability.
\\\textcircled{1} For creation, draw a random integer $j=[(N(b)+1)\times \rm{rndm}]$ ($j=0, 1\cdots N(b)$) to select the time interval to be updated
$(\tau_j^{(b)}, \tau_{j+1}^{(b)})$, where $\tau_j^{(b)}$ with $j=1\cdots N(b)$
is the time position of the $j$th one-spin flip on bond $b$ and
$\tau_0^{(b)}=0, \tau_{N(b)+1}^{(b)}=\beta$.
Draw times for the new one-spin flips using simple uniform probability
density within the selected interval,
$W( \tau_\alpha, \tau_\beta > \tau_\alpha )=const =2/(\tau_{j+1}^{(b)}-\tau_j^{(b)})^2$.
The acceptance ratio for creation is given by
\begin{equation}
R=\min \{1, \frac{(\tau_{j+1}^{(b)}-\tau_j^{(b)})^2}{2}h_x^2\exp\{-\int_{\tau_\alpha}^{\tau_\beta}\Delta U(\tau)d\tau\}\},
\nonumber
\end{equation}
where $\Delta U(\tau)$ is the change of the potential energy (the same notation will be used
for other updates without explicitly mentioning that the integral has to be taken only for
the updated time interval). The proposed configuration change is illustrated in Fig.~\ref{one_spin_flip_creation}.
\begin{figure}[htbp]
\vspace{0cm}
 \includegraphics[width=0.85\columnwidth]{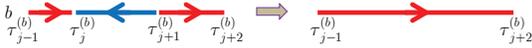}
 \caption{(Color online) Annihilate a pair of one-spin flips. }
 \label{one_spin_flip_annihilation}
\end{figure}
\\\textcircled{2} For annihilation, exit the update if $N(b)<2$;
otherwise draw a random integer $j=[(N(b)-1)\times \rm{rndm}]+1$ ($j=1,2\cdots N(b)-1$) and
propose to annihilate $j$th and $(j+1)$th one-spin flips.
The acceptance ratio is
\begin{equation}
R=\min \{1, \frac{2}{(\tau_{j+2}^{(b)}-\tau_{j-1}^{(b)})^2}\frac{1}{h_x^2}\exp\{-\int \Delta U(\tau)d\tau\}\}\,.
\nonumber
\end{equation}
The update is illustrated in Fig.~\ref{one_spin_flip_annihilation}.
\\[2ex]\textbf{Update 2} Move a one-spin flip.
\\\textbf{Step} 2.1 Randomly select a bond $b$. Exit this update if $N(b)=0$.
\\\textbf{Step} 2.2 Randomly select a one-spin flip, 
$j=[N(b)\times \rm{rndm}]+1$ ($j=1,2\cdots N(b)$).
Let its time position be $t$ and times of previous and next flips on bond $b$ (regardless of their types)
be $t_1$ and $t_2$ respectively, with noting that the worldline is $\beta-$periodic.
\\\textbf{Step} 2.3
\\\textcircled{1} If $t_1< t< t_2$, propose to move the flip to $t'=(t_2-t_1)\times \rm{rndm}$$+t_1$;
\\\textcircled{2} Otherwise, the new time position of the flip can be either in $(0, t_2)$ or $(t_1, \beta)$, i.e.
$t'={\rm mod} \{ (\beta + t_2 - t_1)\times \rm{rndm}+t_1 , \beta \}$.
\\\textbf{Step} 2.4 The acceptance ratio is simply given by the potential energy change
\begin{equation}
R=\min \{1, \exp\{-\int \Delta U(\tau)d\tau\}\}.
\nonumber
\end{equation}
\\[2ex]\textbf{Update 3} Create/Annihilate a pair of four-spin flips.
This update is similar to \textbf{Update 1}.
\\\textbf{Step} 3.1 Randomly select a site $s$. Let the number of
the existing four-spin flips at $s$ is $N(s)$.
\\\textbf{Step} 3.2 With $\sigma_b^z (\tau=0)$ on bonds $b_1, \dots , b_4$
attached to $s$ (see Fig.~\ref{four_spin_flip_creation}) kept unchanged,
propose to create or annihilate a pair of four-spin flips at site $s$ with equal probability.
\\\textcircled{1} In creation, draw a random integer $j=[(N(s)+1)\times \rm{rndm}]$ ($j=0,1\cdots N(s)$)
to identify an interval $(\tau_j^{(s)},\tau_{j+1}^{(s)})$,
where $\tau_j^{(s)}$ is the time position of the $j$th four-spin flip at site $s$
with $\tau_0^{(s)}=0$ and $\tau_{N(s)+1}^{(s)}=\beta$.
Propose time positions for the new four-spin flips from a simple uniform probability density
within the selected interval,
$W( \tau_\alpha, \tau_\beta > \tau_\alpha )=const =2/(\tau_{j+1}^{(s)}-\tau_j^{(s)})^2$.
The acceptance is then
\begin{equation}
R=\min \{1, \frac{(\tau_{j+1}^{(s)}-\tau_j^{(s)})^2}{2}J_s^2\exp\{-\int \Delta U(\tau)d\tau\}\},
\nonumber
\end{equation}
Fig.~\ref{four_spin_flip_creation} provides an illustration of the configuration change
implied by the proposed update.
\\\textcircled{2} In annihilation, exit the update if $N(s)<2$; otherwise
draw a random integer $j=[(N(s)-1)\times \rm{rndm}]+1$ ($j=1,2\cdots N(s)-1$),
and propose to annihilate the $j$th and $(j+1)$th four-spin flips.
The acceptance ratio is
\begin{equation}
R=\min \{1, \frac{2}{(\tau_{j+2}^{(s)}-\tau_{j-1}^{(s)})^2}\frac{1}{J_s^2}\exp\{-\int \Delta U(\tau)d\tau\}\},
\nonumber
\end{equation}
and the configuration changes are illustrated in Fig.~\ref{four_spin_flip_annihilation}.
\\[2ex]\textbf{Update 4} Move a four-spin flip. This update is similar to \textbf{Update 2}.
\\\textbf{Step} 4.1 Select a site $s$ at random. Exit this update if $N(s)=0$.
\\\textbf{Step} 4.2 Randomly select a four-spin flip on site $s$, $j=[N(s)\times \rm{rndm}]+1$ ($j=1,2\cdots N(s)$).
Let its time position be $t$ while $t_1$ ($t_2$) be the time position of the previous (next) flip (regardless of its type)
changing the state of any of the bonds $b_1, ...b_4$ connected to site $s$.
\\\textbf{Step} 4.3
\\\textcircled{1} If $t_1< t< t_2$, propose to move the flip to $t'=(t_2-t_1)\times \rm{rndm}$$+t_1$;
\\\textcircled{2} If $t< t_2\leq t_1$ or $t_2\leq t_1< t$, the new time position of the flip
is $t'={\rm mod} \{ (\beta + t_2 - t_1)\times \rm{rndm}+t_1 , \beta \}$.
\\\textbf{Step} 4.4 The acceptance ratio is
\begin{equation}
R=\min \{1, \exp\{-\int \Delta U(\tau)d\tau\}\} \; .
\nonumber
\end{equation}
\\[2ex]\textbf{Update 5} Create/annihilate combinations of one- and four-spin flips.
\\\textbf{Step} 5.1 Randomly select a site $s$.
\\\textbf{Step} 5.2 With $\alpha_0$ kept unchanged, propose to create or annihilate
a four-spin flip on $s$ with equal probability. 
\\\textcircled{1} In creation, draw a random integer $k=[(N(s)+1)\times \rm{rndm}]$ ($k=0,1\cdots N(s)$). Propose the time position for a new four-spin flip within the interval $(\tau_k^{(s)},\tau_{k+1}^{(s)})$ using $t=(\tau_{k+1}^{(s)}-\tau_k^{(s)})\times \rm{rndm}$$+\tau_k^{(s)}$. The detailed-balance factor associated with this proposal is $f_c(s)=(N(s)+1)(\tau_{k+1}^{(s)}-\tau_k^{(s)})J_s$.
\\\textcircled{2} In annihilation, exit the update if $N(s)=0$; otherwise draw a random integer $k=[N(s)\times \rm{rndm}]+1$ ($k=1,2\cdots N(s)$),
and propose to annihilate the $k$th four-spin flip, of which the time position is at $t$.
The detailed-balance factor associated with this proposal is $f_a(s)=N(s)$.
Note that in terms of the current configuration parameters $r(s)=f_c(s)/f_a(s)=(\tau_{k+1}^{(s)}-\tau_k^{(s)})J_s$
in creation and $r(s)=f_a/f_c=1/((\tau_{k+1}^{(s)}-\tau_{k-1}^{(s)})J_s)$ in annihilation.
\\\textbf{Step} 5.3 On bond $b_i$ ($i$=1, 2, 3, 4) connected to site $s$, the time $t$ falls within some
time interval ($\tau_j^{(b_i)}$, $\tau_{j+1}^{(b_i)}$), where $t$ is the time position for
either the created or the annihilated four-spin flip in \textbf{Step} 5.2. For each bond $b_i$, propose
to create or annihilate a one-spin flip on this interval.
[As before, $\alpha_0$ is kept unchanged, and the probabilities for creation and annihilation are set equal.]
\\\textcircled{1} For creating a one-spin flip, select new time variable at
random $t^{(b_i)}=(\tau_{j+1}^{(b_i)}-\tau_j^{(b_i)})\times \rm{rndm}$$+\tau_j^{(b_i)}$.
The corresponding detailed-balance factor is $f_c(b_i)=(\tau_{j+1}^{(b_i)}-\tau_j^{(b_i)})h_x$.
\\\textcircled{2} For annihilating a one-spin flip, randomly choose between labels $j$ and $j+1$ and propose to
annihilate the selected $j'$th one-spin flip (exit the update if $j'=0$ or $N(b_i)+1$).
The detailed-balance factor is $f_a(b_i)=2$.  Again, in terms of the current configuration parameters
$r(b_i)=f_c(b_i)/f_a(b_i)=(\tau_{j+1}^{(b_i)}-\tau_j^{(b_i)})h_x/2$ in creation
and $r(b_i)=2/((\tau_{j'+1}^{(b_i)}-\tau_{j'-1}^{(b_i)})h_x)$ in annihilation.
\\\textbf{Step} 5.4 The acceptance ratio is
\begin{equation}
R=\min \{1, r(s)\prod_{i=1}^{4}r(b_i) \exp\{-\int\Delta U(\tau)d\tau\}\} \; , \nonumber
\end{equation}
Fig.~\ref{kink} provides an illustration of configuration changes
produced in these updates.

\section{Phase Diagram}
We restrict our study of Hamiltonian (\ref{H}) to the parameter subspace
defined by $J_s=J_p$ (and use $J_s$ as a unit of energy).
We also set $\beta = L$ in the Monte Carlo simulation aimed at the study
of ground state properties, where $L\times L =N$ is the square lattice size. For $J_s=J_p$ the location of the self-duality line is given simply by
$h_x=h_z$. Thus, in the following we only need to focus on the region $h_x\leq h_z$ of the phase diagram. The
$h_x$=0 case was studied in Ref.~\onlinecite{Trebst07}
where the authors have shown that TCM with $h_z\ne 0$ is equivalent to the 2D
transverse-field Ising model featuring a continuous phase transition with the dynamic critical exponent $z=1$. Using high-accuracy results for the transverse-field
Ising model\cite{Blote02}, we immediately establish the critical value
of the $h_z$-field, $h_z^{(c)}(h_x=0)$=0.328474(3). When $h_x$ is finite but weak,
TCM undergoes the same Ising-type continuous transition driven
by large $h_z$-field~\cite{Tupitsyn10}.
Though in finite fields the predominantly charge-condensed and predominantly
vortex condensed phases can be continuously transformed into each other, along the self-duality line ($h_x=h_z$) the symmetry between charges and vortices
can be broken in a first-order transition over a certain range of fields $h_z$ (=$h_x$).

\begin{figure}[htbp]
\vspace{0cm}
 \includegraphics[width=0.8\columnwidth]{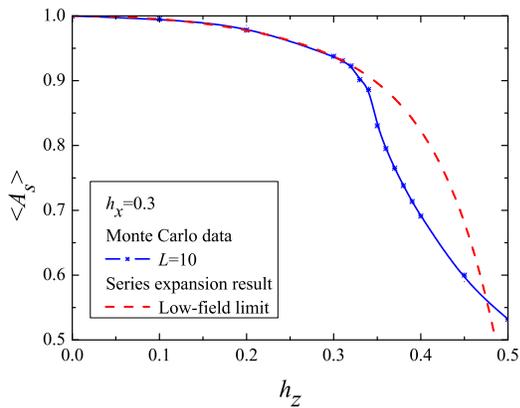}
 \caption{(Color online) $\langle A_s\rangle$ versus $h_z$ for $h_x=0.3$.}
 \label{As}
\end{figure}

\begin{figure}[htbp]
\vspace{0cm}
 \includegraphics[width=0.8\columnwidth]{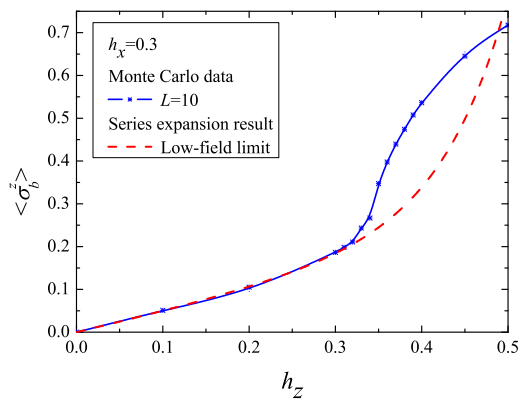}
 \caption{(Color online) $\langle\sigma_b^z\rangle$ versus $h_z$ for $h_x$=0.3.}
 \label{sigma_b^z}
\end{figure}

\emph{The second-order phase transition line.} In a finite $h_x$ field, our Monte Carlo data show that both $\langle A_s\rangle$ (Fig.~\ref{As}) and $\langle\sigma_b^z\rangle$ (Fig.~\ref{sigma_b^z})
develop strong non-analytic features at the transition point indicating the onset of charge condensation. To compare our data to the series expansion results, we also compute $\langle A_s\rangle$ and $\langle\sigma_b^z\rangle$ using the Hellmann-Feynman theorem for the ground-state energy at order 10 in the low-field limit \cite{Dusuel11}.
As can be seen in Fig.~\ref{As} and Fig.~\ref{sigma_b^z}, the results of the two approaches match very well at low fields, but the series expansion result is completely missing the non-analytic behavior in larger fields.

\begin{figure}[htbp]
\vspace{0cm}
 \includegraphics[width=0.8\columnwidth]{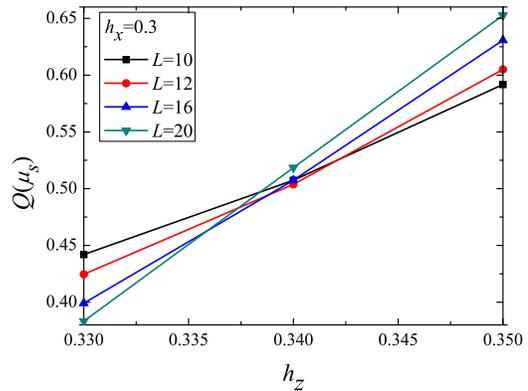}
 \caption{(Color online) Binder cumulant $Q(\mu_s)$ for linear system sizes $L=10,12,16,20$ versus $h_z$
 for $h_x$=0.3.}
 \label{Q_ms}
\end{figure}

There is no explicit local order parameter characterizing the phase transition in terms of original spins
\cite{Vidal09, Tupitsyn10}. However, local (in space)
order parameter can still be found by introducing auxiliary variables and applying gauge transformation to the Hamiltonian (\ref{H})\cite{Tupitsyn10}.
As an effective magnetic moment associated with
operator $A_s$, we introduce an order parameter $\mu_s$ which can be easily sampled
in the Monte Carlo simulation:
\begin{eqnarray}
  &\mu_s=\pm\frac{1}{\beta}\{(\tau_1^{(s)}-0) - (\tau_2^{(s)}-\tau_1^{(s)}) + \cdots + (-1)^{N(s)-1} &\nonumber \\
  &(\tau_{N(s)}^{(s)}-\tau_{N(s)-1}^{(s)})+(-1)^{N(s)}(\beta-\tau_{N(s)}^{(s)})\},&
\end{eqnarray}
where $\tau_1^{(s)}<\tau_2^{(s)}<\cdots<\tau_{N(s)}^{(s)}$ are the imaginary
time positions of four-spin flips on site $s$. Since the sign of $\mu_s$
is arbitrary, we define its second and fourth moments as:
\begin{equation}
\langle\mu_s^2\rangle=\langle\sum_s\mu_s^2\rangle/N, \langle\mu_s^4\rangle=\langle\sum_s\mu_s^4\rangle/N \,.
\end{equation}
The Binder cumulant~\cite{Binder81} of $\mu_s$ is defined as
\begin{equation}
Q(\mu_s)=\langle\mu_s^2\rangle^2/\langle\mu_s^4\rangle \, .
\end{equation}
As shown in Fig.~\ref{Q_ms}, $Q(\mu_s)$ behaves linearly in the vicinity of the critical point.
The finite-size behavior of $Q(\mu_s)$ near the critical point can be parameterized as \cite{Blote95},
\begin{eqnarray}
  &Q(\mu_s)=Q+\sum_{k=1} a_k(h_z-h_z^{(c)})^k L^{k y_t} &\nonumber
\\&+b_1L^{y_i}+b_2L^{y_2}+c_1L^{y_3}(h_z-h_z^{(c)})+\cdots \; , &
 \label{scaling}
\end{eqnarray}
which includes both the leading scaling terms and corrections to scaling.
The Monte Carlo data are fitted on the basis of this formula, according to a least-squared criterion.
The fitting result shows that the critical point for $h_x=0.3$ is located at  $h_z^{(c)}(h_x=0.3)=0.333(1)$.
Our limited system sizes are insufficient for accurate determination of the critical exponent $y_t$,
which should be 1.5868(3)~\cite{Deng03} since the transition is supposed to be of 3D Ising universality.


\begin{figure}[htbp]
  \includegraphics[width=0.8\columnwidth]{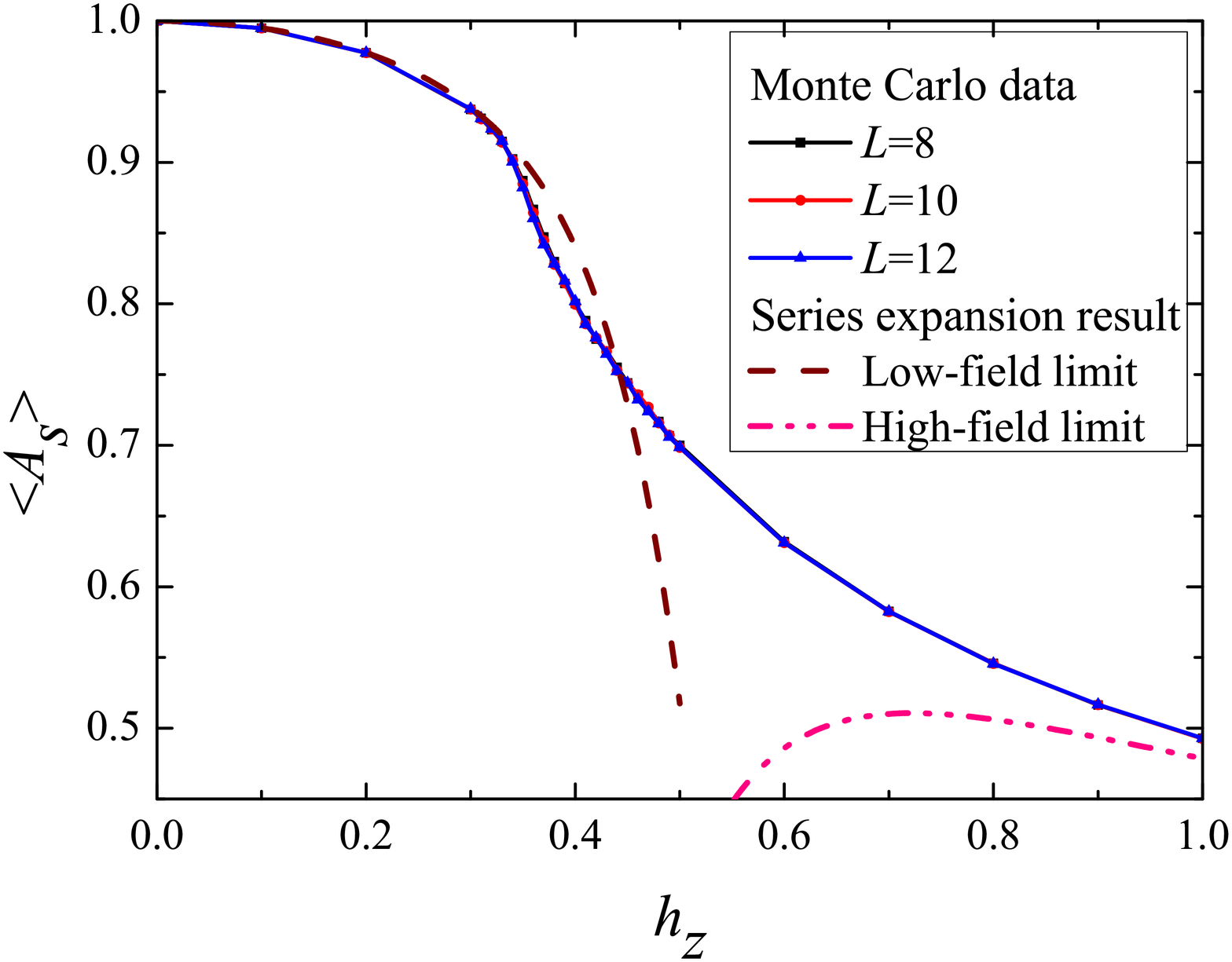}
  \caption{(Color online) $\langle A_s\rangle$ versus $h_z$ along the self-duality line.}
  \label{A_x_z}
\end{figure}

\begin{figure}[htbp]
  \includegraphics[width=1\columnwidth]{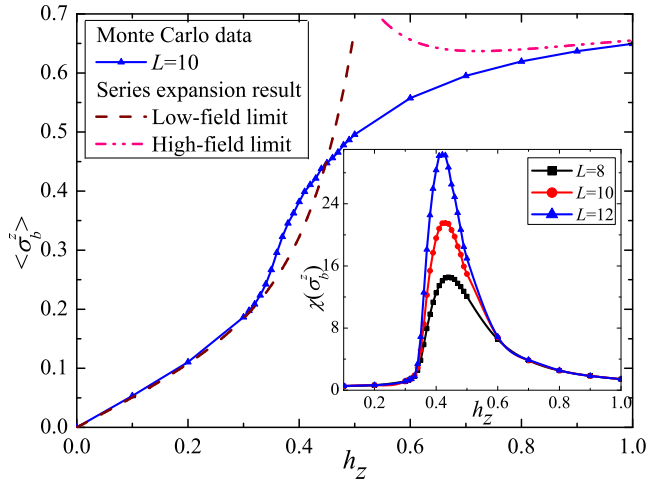}
  \caption{(Color online) $\langle\sigma_b^z\rangle$ versus $h_z$ along the self-duality line. The inset shows the susceptibility of $\sigma_b^z$
  defined as $\chi(\sigma_b^z)=2L^2\beta(\langle(\sigma_b^z)^2\rangle-\langle\sigma_b^z\rangle^2)$.}
  \label{s_x_z}
\end{figure}

\emph{The first-order phase transition line.}
Our Monte Carlo data for $\langle A_s\rangle$ and $\langle\sigma_b^z\rangle$ along the self-duality line are respectively shown in Fig.~\ref{A_x_z} and Fig.~\ref{s_x_z}, which also display the series expansion results both at order 10 in the low-field limit \cite{Dusuel11} and at order 5 in the high-field limit \cite{Vidal09}. Again, the Monte Carlo data and the series expansion result are in good agreement at low fields (approximately $h_z<0.34$). However, the series expansion results do not provide an accurate description
of the model in the parameter region $0.4<h_z<0.8$. As shown in Fig~\ref{s_x_z}, the susceptibility grows as the lattice sizes increases at intermediate-field region, which is a sign of phase transitions.

\begin{figure}
\vspace{0cm}
 \includegraphics[width=1\columnwidth]{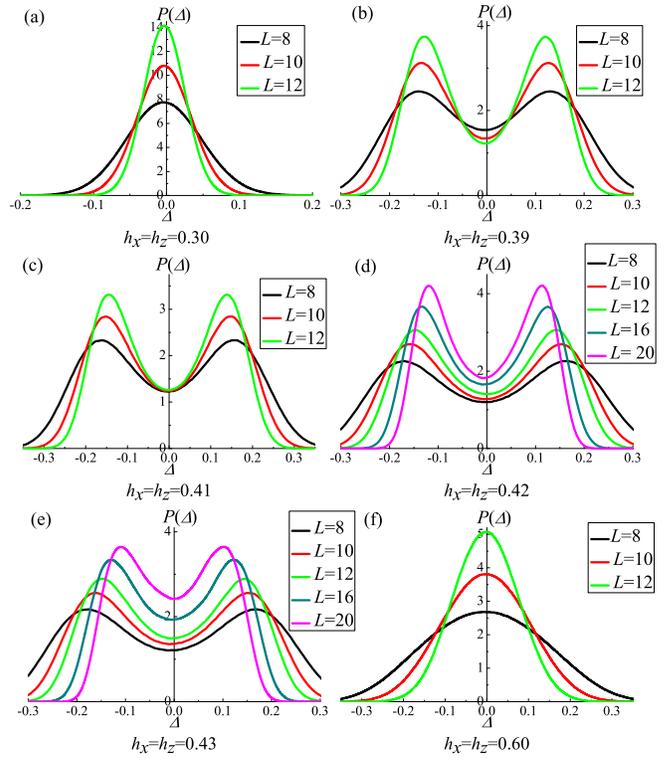}
 \caption{(Color online) Probability distribution of $\Delta$ along the self-duality line. $P(\Delta)$ with bimodal structure are reweighted so that the two peaks have equal heights.}
 \label{dis}
\end{figure}

Along the self-duality line, there is an obvious symmetry between charges and vortices. To see if the symmetry is broken we sample the probability distribution $P(\Delta )$
for the quantity $\Delta=A_s-B_p$.
By symmetry $P(\Delta)$ is supposed to be an even function of $\Delta$.
The evolution of $P(\Delta)$ along the self-duality line is shown in Fig.~\ref{dis}.
We observe that $P(\Delta)$ at $h_x=h_z$=0.3 and 0.6 has a single Gaussian-type peak
centered at $\Delta =0$, while at $h_x=h_z=0.39$ $P(\Delta)$ has two peaks
at finite values of $|\Delta |$.
The double-peak feature of $P(\Delta)$ is a smoking-gun evidence
for the symmetry breaking between charges and vortices in the first-order transition.
At $h_x=h_z=0.42$ and 0.43, the bimodal structure does appear for the available lattice sizes,
but the local minima around $\Delta=0$ increases with system size and thus should vanish for sufficiently large sizes.
From finite-size scaling theory, this means that despite of the absence of symmetry breaking, these two points
are close to the 1st-order phase transition.
Together with the histogram for $h_x=h_z=0.41$, it is conservative to expect that the end point of the
1st-order transition line lies in range $(0.41,0.42]$.
\begin{figure}[htbp]
  \includegraphics[width=0.8\columnwidth]{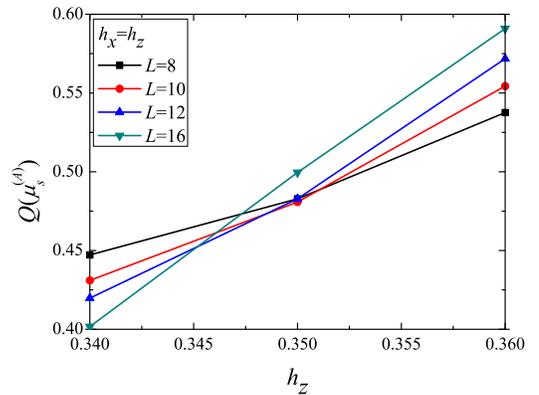}
  \caption{(Color online) $Q(\mu_s^{(A)})$ versus $h_z$ along the self-duality line.}
  \label{Q_ms_A_x_z}
\end{figure}
\begin{figure}[htbp]
  \includegraphics[width=0.8\columnwidth]{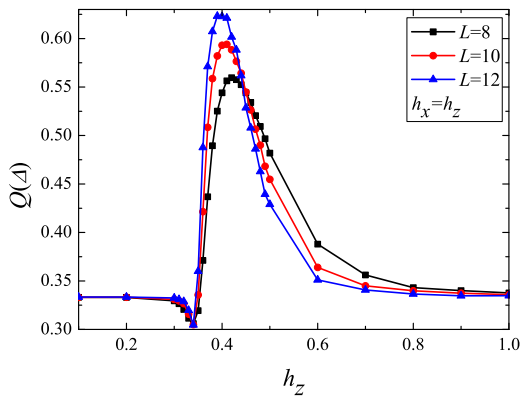}
  \caption{(Color online) $Q(\Delta)$ versus $h_z$ along the self-duality line.}
  \label{Q_d_x_z}
\end{figure}
\begin{figure}[htbp]
  \includegraphics[width=0.8\columnwidth]{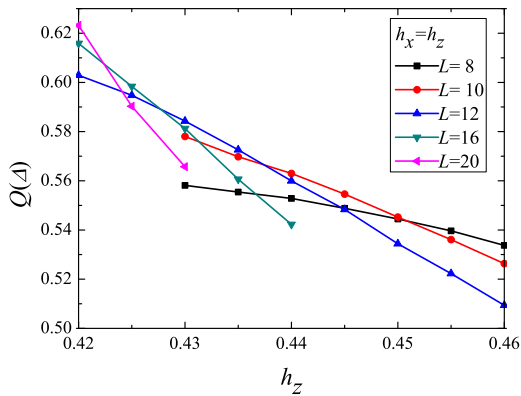}
  \caption{(Color online) $Q (\Delta)$ versus $h_z$ near the end point of the 1st-order transition line.}
  \label{Q_d_0.4}
\end{figure}
\begin{figure}[htbp]
  \includegraphics[width=0.8\columnwidth]{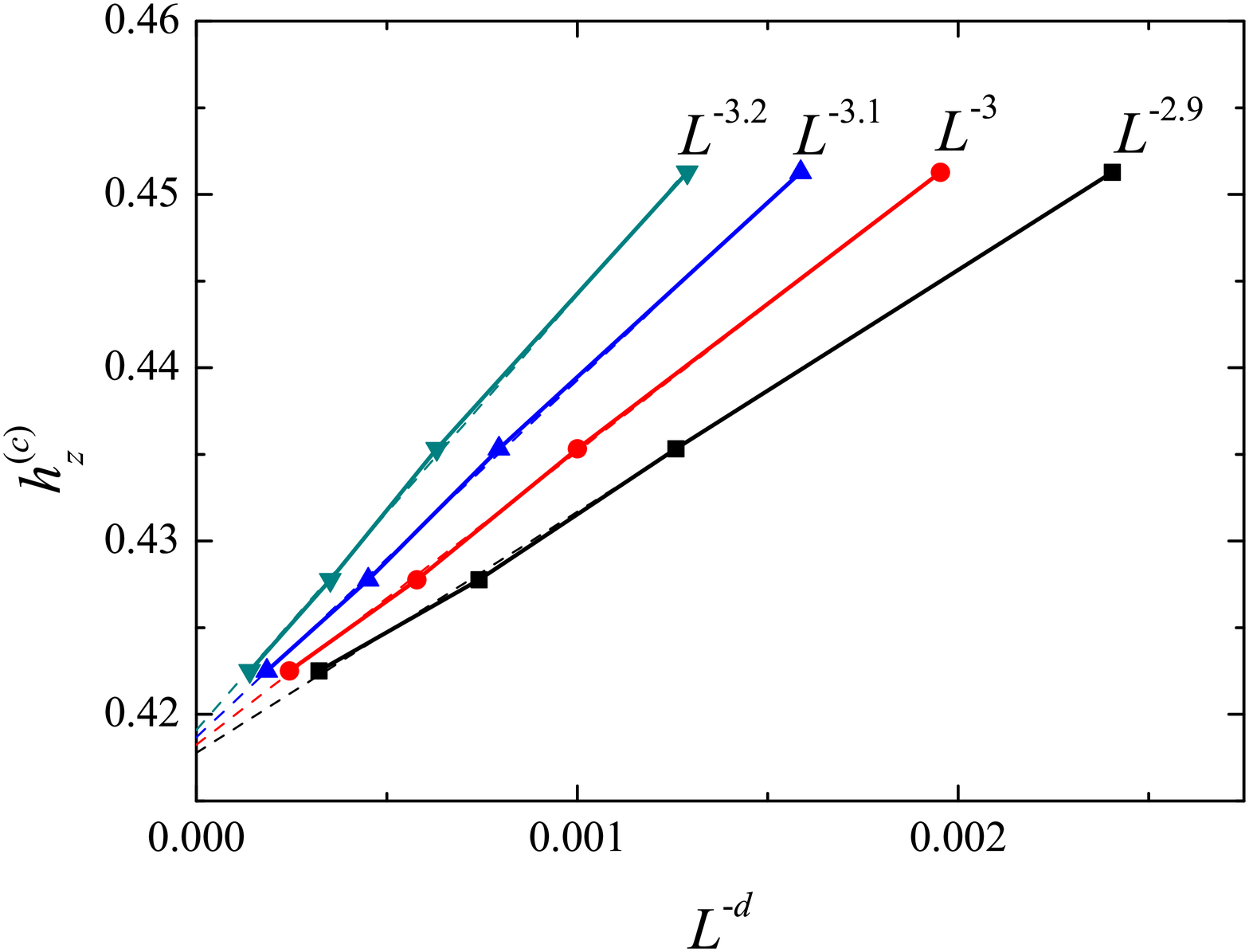}
  \caption{(Color online) Extrapolation of intersection points for $Q(\Delta,L)$ curves for different system sizes.}
  \label{Q_inf}
\end{figure}

Symmetry breaking between charges and vortices leads to the 2-fold degenerate ground state, which can be classified respectively by the predominant condensation of charges ($\Delta=A_s-B_p<0$) and predominant condensation of vortices ($\Delta=A_s-B_p>0$).
Since $\mu_s$ is used as the order parameter for the phase transition to the charge-condensed phase we also sample it near the low-field end of the first-order transition line
but only in configurations with $\Delta=A_s-B_p<0$. We use $\mu_s^{(A)}$ notation to stress that $\mu_s$ was sampled selectively this way.
The behavior of the Binder cumulant $Q(\mu_s^{(A)})$ near the point where all three transition lines meet is shown in Fig.~\ref{Q_ms_A_x_z}.
The Monte Carlo data for $Q(\mu_s^{(A)})$ for different system sizes were also fitted on the basis of Eq.~(\ref{scaling}), showing that the first-order transition line starts around $h_x^{(c)}=h_z^{(c)}=0.340(2)$, in good quantitative agreement with the result $h_x^{(c)}=h_z^{(c)}=0.3406(4)$ obtained in Ref.~\onlinecite{Vidal09}. The agreement is expected since the two approaches give almost the same result for $\langle A_s\rangle$ and $\langle\sigma_b^z\rangle$ in low field.

The Binder cumulant of $\Delta$ along the self-duality line is shown in Fig.~\ref{Q_d_x_z}.
Though $Q(\Delta)$ is highly non-linear near the low-field critical point it
behaves linearly near the high-field end point of the first-order transition line.
Near this end point, the analysis of intersections of $Q(\Delta)$ for different system sizes
provides an accurate estimate of the critical point (Fig.~\ref{Q_d_0.4}).
The extrapolation of intersection points for two subsequent sizes
is shown in Fig.~\ref{Q_inf} using power-law variable $L^{-d}$ for the horizontal axis.
Extrapolations for $L^{-2.9}, L^{-3}, L^{-3.1}$ and $L^{-3.2}$ show that the first-order transition line
ends at $h_x^{(c)}=h_z^{(c)}=0.418(2)$.
This is consistent with our expectation based on histograms, 
and differs from the result $h_x^{(c)}=h_z^{(c)}=0.48(2)$ of the series expansion
\cite{Vidal09} which is questionable in this parameter range. It is also possible that part of the discrepancy is due to
systematic error introduced by Dlog Pad\'{e} approximation/extrapolation, used in Ref.~\onlinecite{Vidal09}.
Series expansion to higher order about the high field limit and the use of other extrapolation schemes,
can provide estimates for the end point in agreement with our value \cite{Schmidt12}.

\begin{figure}[htbp]
  \includegraphics[width=0.8\columnwidth]{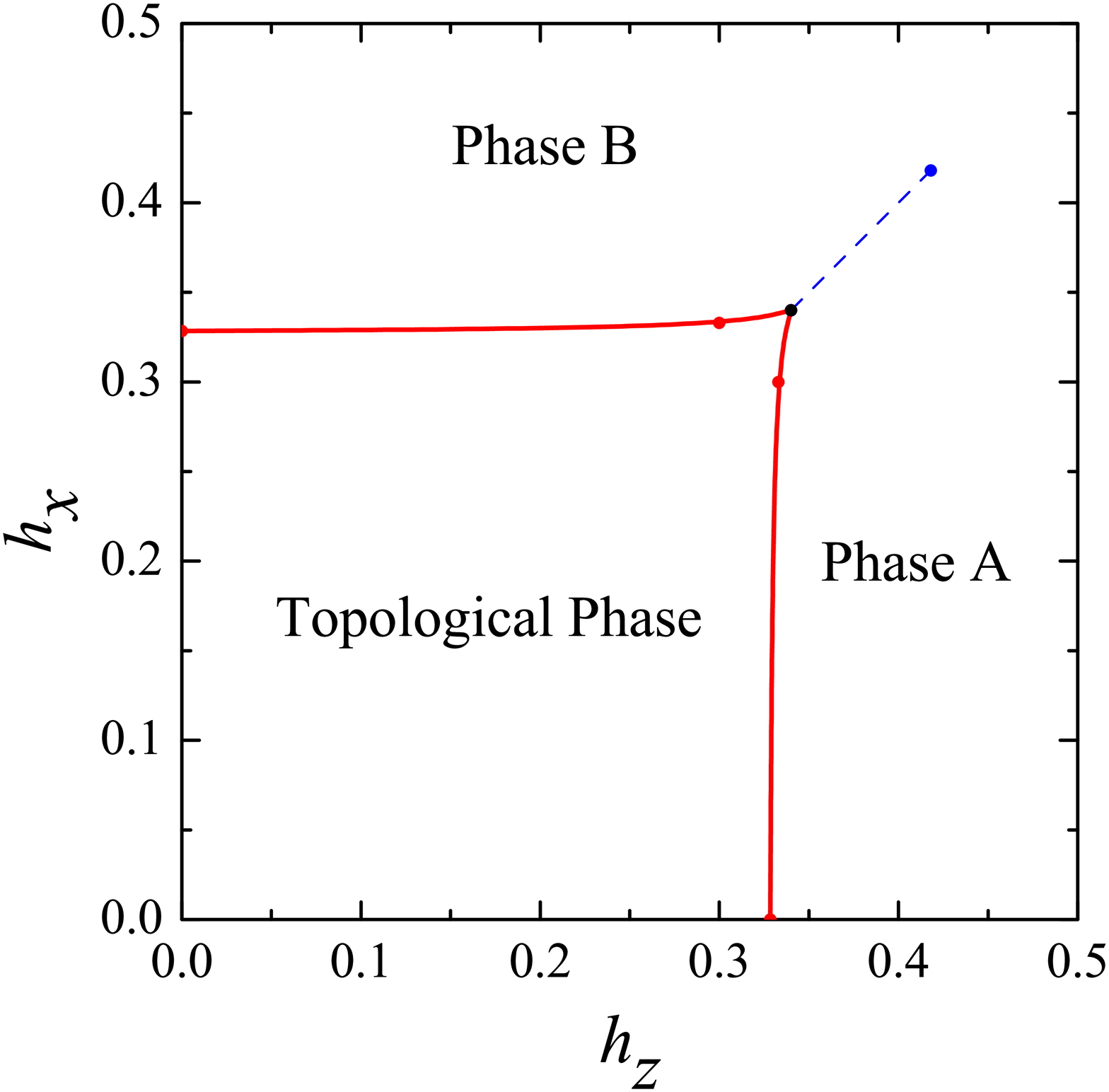}
  \caption{(Color online) Phase diagram of the toric code model in magnetic field. The second-order transition lines are shown by full (red) lines and the
  first-order transition line is represented by dashed (blue) line.}
  \label{phase_diagram}
\end{figure}

The resulting phase diagram of TCM in magnetic field is shown in Fig.~\ref{phase_diagram}.
The continuous phase transition lines separate topologically protected ground states
from magnetically ordered phases A and B characterized by condensation of charges
and vortices, respectively \cite{Trebst07, Tupitsyn10}.
Along the blue dashed line in Fig.~\ref{phase_diagram}, the symmetry between charges and vortices
is broken in the first-order transition which is accompanied by an
abrupt change in the density of charges and vortices.
However, when large enough external field is continuously rotated in the ($x, z$) plane one can continuously
convert phases A and B into each other without inducing a phase transition.

\section{Summary}
To simulate the toric code model in a parallel field, we developed an algorithm based on the continuous-time Monte Carlo approach. It allows one to perform simulations of quantum spin models with kinetic/exchange terms involving four spins. The phase diagram was constructed
using finite-size scaling analysis of Binder cumulants. The two second-order transition lines enclosing the topological phase and related by self-duality were found to be in excellent
quantitative agreement with the study performed in Ref.~\onlinecite{Vidal09}. As for the end point of the first-order transition line, our numerical result provides a potentially better estimate, compared to the result obtained in Ref.~\onlinecite{Vidal09}.
System size limitations did not allow us to get enough resolution to answer the question of how the three phase
transition lines merge around $h_z=h_x=0.34$ \cite{Tupitsyn10}, and, in particular,
to verify that they all terminate at the same point. This longstanding problem requires
further investigation.

Recently, the toric code model has been generalized to $\mathbb{Z}_N$ degrees of freedom \cite{Schulz11}. The Monte Carlo algorithm developed in this paper can be used to study the $\mathbb{Z}_3$ toric code model in a parallel field by adjusting the number of flips to be changed in each elementary Monte Carlo update. This is an interesting problem to look at.

\section{Acknowledgements}
We thank Boris Svistunov and Jianping Lv for helpful discussions.
This work was supported by NSF Grant PHY-1005543, grant from the Army Research Office
with funding from the DARPA OLE program, NSF of China under Grant No. 10975127,
CAS, and the National Fundamental Research Program under Grant No. 2011CB921300.

\end{document}